\documentstyle[aps,preprint,amsfonts]{revtex}
\begin{document}
\draft
\title{Quantum Coding Theorem for Mixed States}
\author{Hoi-Kwong Lo}
\address{
 School of Natural Sciences, Institute for Advanced Study, Olden Lane,\\
 Princeton, NJ 08540, U.S.A.
}
\author{and}
\address{
Hong Kong Telecom Institute of Information Technology, \\ Hong
Kong University of Science and Technology, \\ Clear Water Bay, Hong Kong}
\date{\today}
\preprint{IASSNS-HEP-95/23}
\maketitle
\mediumtext
\begin{abstract}
We prove a theorem for coding mixed-state
quantum signals. For a class of
coding schemes, the von Neumann entropy $S$
of the density operator describing an ensemble of mixed quantum signal
states is shown to be equal to the number of spin-$1/2$ systems necessary to
represent the signal faithfully.
This generalizes previous works on coding
pure quantum signal states and is analogous to the Shannon's noiseless
coding theorem of classical information theory.
We also discuss an example of a more general class of coding schemes which
{\em beat} the limit set by our theorem.
\end{abstract}
\medskip
A key concept in classical information theory developed
by Shannon\cite{Sha} and others\cite{Cover} is the entropy.
For a discrete random variable (source) $A$, it is defined by
\begin{equation}
H(A)= - \sum_a p(a) \log_2 p(a). \label{centropy}
\end{equation}
Coding is an important issue in information theory. In particular,
one may be interested in
representing the messages produced by the source $A$ by a sequence
of binary digits (bits) as short as possible. Suppose that
$A$ emits a seqence of independent messages. If we allow
ourselves to code entire blocks of independent messages together and
tolerate an arbitrarily small error in the signals reconstructed from the coded
version, it turns out that the mean number of bits per message
needed can be arbitrarily made close to $H(A)$.

Recently, there
has been much interest in the subject of quantum computation.
Current investigations\cite{qc} include the physical implementation of
quantum computers, quantum complexity theory, quantum teleportation and
quantum coding. In quantum coding,
Schumacher\cite{Sch} and Jozsa and Schumacher\cite{JozSch}
have considered the possibility that the signals are pure
quantum states which are not necessarily orthogonal to one another.
Suppose that a quantum source $A$ emits a sequence of independent signals,
each of which is a pure state from the list $|a_1 \rangle, \cdots, |a_m
\rangle$
occuring with probabilities $p_1, \cdots , p_m$. We may associate
the density matrix
\begin{equation}
\rho = \sum_{i=1}^m p_i | a_i \rangle \langle a_i |  \label{qentropy}
\end{equation}
to the source.
By analogy with the classical measure of information, the bit, as a
2-state classical system, Schumacher used the term ``qubit'' (meaning quantum
bit) for
the quantum state storage capacity of a two-dimensional Hilbert space.
Note that, unlike a classical bit which can only
take on a value of either $0$ or $1$, the state of a qubit can be
in some coherent superposition of $0$ and $1$. i.e. the state of
a qubit $|u \rangle = \alpha |0 \rangle + \beta |1 \rangle $ where
$\alpha , \beta \in {\Bbb C}$ and $|\alpha|^2 +|\beta|^2 = 1$.
Moreover, a qubit is capable of being entangled with the states of other
qubits. For example, the state
${1 \over \sqrt 2} \left( |10 \rangle - |01 \rangle \right)$
is allowed.
The polarization of a single photon, for example, has a storage capacity
of one qubit. We wish to encode the signals with a least possible
number of Hilbert space dimensions. Once again, block coding may be used
and a small error may be allowed. In other words, we consider
a $K$-blocked version $A_K$ of $A$. If $A$ has $m$ distinct signal states
in a Hilbert space $H_n$ (of dimension $n$), then $A_K$ has $m^K$ signals
in $H_{n^K}$ (of dimension $n^K$). In order to code the signals
with a minimum number of Hilbert space dimensions, typically part of
a system will be discarded during the coding.
Therefore, the signal $|a_i \rangle$ is reconstituted as a mixed state
with density matrix $W_i$.
In Refs. \cite{Sch} and \cite{JozSch} the concept of fidelity
\begin{equation}
F= \sum_i p_i \langle a_i | W_i |a_i \rangle \label{fidelity}
\end{equation}
was introduced. Notice that $\langle a_i | W_i |a_i \rangle $ is the
probability that the state $W_i$ passes the yes/no test of being
the state $a_i$. $0 \leq F \leq 1$ is the average probability of
passing the test.

Analogous to the classical information theory, we introduce
the von Neumann entropy
\begin{equation}
S(\rho)= - {\text Tr} \rho \log_2 \rho. \label{Neumann}
\end{equation}
The quantum noiseless coding theorem for {\em pure} states
proved in Refs. \cite{Sch} and \cite{JozSch} states the following.
Given any quantum source with von Neumann entropy $S(\rho)$ and
any $\epsilon , \delta > 0$,

(a) If $S(\rho)+ \delta$ qubits are available per signal, then
for each sufficiently large $N$, there exists a coding scheme with
fidelity $F> 1 - \epsilon$ for signal strings of length $N$.

(b) If $S(\rho)- \delta$ qubits are available per signal, then any
coding scheme for strings of length $N$ will have a fidelity
$F < \epsilon$ for all sufficiently large $N$.

Therefore, the von Neumann entropy may be interpreted
as the minimal number
of qubits needed for reliable (almost noiseless) coding.
This noiseless coding theorem works only for {\em  pure} signal
states. It is natural to generalize it to consider signals
which are {\em mixed} states $\Pi_a$,
with $\rho =\sum_a p(a) \Pi_a$.
As noted in Refs. \cite{Sch} and \cite{JozSch},
it is not clear how to proceed. A naive generalization of
the fidelity,
\begin{equation}
F= \sum_a p(a) {\text Tr} \Pi_a W_a \label{fidelityn}
\end{equation}
is not close to unity even when $W_a = \Pi_a$ for all signals.
To quantify the amount of distortion of a particular coding scheme,
a notion of the distance between two mixed states is desired.
Such a concept has been introduced by Anandan\cite{Ana}
in the study of geometric phases. Let $\cal D$ denote
the set of density operators representing the states of
a given quantum system. ($\cal D$ consists of the set of
Hermitian operators in the Hilbert space of this system
with nonnegative eigenvalues and trace equal to 1.)
$\cal D$ is a topological space with the pure states contained in its
boundary. The set of pure states can be identified with the projective
Hilbert space $\cal P$.
The inner product structure of
a Hilbert space naturally induces a metric, namely the Fubini-Study metric
on the projective Hilbert space $\cal P$
which can be extended into the rest of
$\cal D$. More concretely, the distance between two points $p$ and $p'$
in $\cal P$ is defined by
\begin{equation}
s(p, p')^2 =4 (1 - | \langle \psi | \psi' \rangle |^2 ) \label{metricp}
\end{equation}
where $| \psi \rangle $ and $| \psi' \rangle$ are two normalized states
contained in $p$ and $p'$. It is simple to check that $s(p,p')$ satisfies
all the axioms for a metric. Suppose that $p$ and $p'$ are separated
by an infinitesimal distance $ds$ in $\cal P$:
\begin{equation}
ds^2 =4 (1 - | \langle \psi | \psi' \rangle |^2 )= 2 {\text Tr} (
\rho - \rho')^2  \label{metricpn}
\end{equation}
where the last equality follows from ${\text Tr} (\rho^2) = {\text Tr}
(\rho'^2) =1 $.
This defines a Riemannian metric on $\cal P$, called the Fubini-Study
metric. It is therefore reasonable to introduce a flat metric
\begin{equation}
dS^2 =2 {\text Tr} (d \rho^2) \label{metricd}
\end{equation}
on $\cal D$. When restricted to the
pure states, it becomes the Fubini-Study metric.

Suppose that a quantum source produces a sequence of signals, each of which
is a mixed state from the list $\Pi_1, \cdots \Pi_m$, with probabilities
$p_1, \cdots, p_m$ and that after coding, the signal $\Pi_a$ is reconstituted
as $W_a$.
Motivated by the above discussion, we define
the distortion
\begin{equation}
D= \sum_a p_a {\text Tr } (\Pi_a - W_a)^2 . \label{distort}
\end{equation}
Notice that $ 0 \leq D \leq 2$ and $D= 0$ if and only if $\Pi_a = W_a$.
This definition is reasonable because
$\Pi_a - W_a$ is the deviation of $\Pi_a$ from $ W_a$. To obtain
a real-valued function, we take the trace. However, ${\text Tr } (\Pi_a - W_a)$
is identically zero. It is, therefore, natural to consider
${\text Tr } (\Pi_a - W_a)^2$ and take the ensemble average.

The ensemble of signals emitted by the source can be represented
by the density operator
\begin{equation}
\rho= \sum_a p_a \Pi_a . \label{Neumannn}
\end{equation}

Consider the following
communication scheme discussed by Schumacher\cite{Sch}.
Suppose that the signal is represented by a system
$X$ which is composed of two subsystems, $C$ (for
``channel'')
and $E$ (for ``extra''). Only the channel subsystem $C$ is transmitted to the
receiver and the subsystem $E$ is simply discarded.
To recover (some approximation of) the signal, we add to
the channel system an auxillary
system $E'$ that is a copy of the discarded extra system $E$.
Schumacher called such a communication scheme an {\em approximate transposition
via the limited channel} $C$.
For this type of communciation schemes, we have the following theorem:

{\em Quantum Noiseless
Coding Theorem for Mixed States.} For any quantum source which produces
mixed signal states $\Pi_a$'s with probabilities $p_a$'s, define the
von Neumann
entropy $S(\rho)$ as in Eq.\ (\ref{qentropy}).
For any $\epsilon , \delta >0 $,

(a) if $S(\rho) + \delta$ qubits are available per signal, then for
each sufficiently large $N$, there exists a coding scheme
with $ D < \epsilon $.

(b) if $S(\rho) - \delta$ qubits are available per signal, then for a
sufficiently large $N$,
any approximate
transposition coding scheme
for a string of length $N$
has a distortion $D \geq \sum_a p_a {\text Tr} \Pi_a^2 - \epsilon$.

This implies that for a given quantum source, $D$ will not tend
to zero unless at least $S(\rho)$ qubits are available per signal.
Therefore, $S( \rho)$ may again be interpreted as the mean number
of bits needed for the noiseless coding of a source which emits signals
that are mixed states if an approximate transposition coding scheme is
used.

To minimize our usage of resources, we would like to code signals on
a $d$-dimensional subspace $\Lambda$ of ${\cal H}_n$. (In
applying the following lemmas to prove the main theorem, we will
use block coding. The signal states will therefore be $K$-blocks
of signals.) Let $|b_1 \rangle
, |b_2 \rangle , \cdots , |b_d \rangle $ be a basis of $\Lambda$
and $|b_{d+1} \rangle
, |b_{d+2} \rangle , \cdots , |b_n \rangle $ a basis of $\Lambda^\perp$,
the orthogonal complement of $\Lambda$.
For each a, $\Pi_a$ can be diagonalized
and expressed in terms of its eigenvectors $|a_i \rangle$ as
\begin{equation}
\Pi_a = \sum_{a_i} q_{a_i} |a_i \rangle \langle a_i | = \sum_{a_i}
 q_{a_i} \Pi_{a_i}
\label{pai}
\end{equation}
where $\Pi_{a_i}= |a_i \rangle \langle a_i |$ and for
each $a$, $\sum_{a_i} q_{a_i} =1$.
Suppose that, with respect to the basis $|b_1 \rangle
, |b_2 \rangle , \cdots , |b_n \rangle $,
\begin{equation}
\Pi_{a_i}=
|a_i \rangle \langle a_i | =\left(
\begin{array}{cc}
M_{a_i} & A_{a_i}^{\dagger} \\
A_{a_i} & N_{a_i }
\end{array}
\right)  \label{piai}
\end{equation}
where $M_{a_i}$ is a $d \times d$ matrix.
We now introduce
an explicit coding scheme based on $\Lambda$. Let $| 0 \rangle$ be an
arbitrary state in $\Lambda$ and $P$ the projection into $\Lambda$.
For each $ \Pi_{a_i}=
|a_i \rangle \langle a_i | $, we measure the observable $P$ on $|a_i \rangle$.
If the result $0$ is obtained, then $| 0 \rangle$ is substituted for the
post measurement state.
In other words, we associate with each $\Pi_{a_i}$ a density
matrix
\begin{equation}
W_{a_i}= \left(
\begin{array}{cc}
M_{a_i} & 0 \\
0  & 0
\end{array}
\right) + (1 - {\text Tr} M_{a_i}) | 0 \rangle \langle 0 | . \label{wai}
\end{equation}

{\em Lemma 1.} Suppose that the sum of the $d$ largest eigenvalues
of the density operator $\rho$ is greater than $1 - \xi$. Let $\Lambda$
be the span of the $d$ eigenvectors of $\rho$ corresponding to the
$d$ largest eigenvalues.
Then the
association $\Pi_a = \sum_{a_i} q_{a_i} \Pi_{a_i}  \longleftrightarrow W_a
= \sum_{a_i} q_{a_i} W_{a_i} ,$ defined by Eq.\ (\ref{wai}) has distortion
$D < 2 \xi$.

{\em Proof:} Note that $f(X)= {\text Tr} (X^2)$ is a convex function.
We have ${\text Tr} [ E(X)]^2 \leq E ({\text Tr} X^2 ) $ where $E(X)$ denotes
the weighted mean of a variable $X$.
Denoting $\Pi_a - W_a$ by $X_a$ and $\Pi_{a_i} -W_{a_i}$ by $X_{a_i}$,
the distortion
\begin{eqnarray}
D &=& \sum_a p_a {\text Tr} (X_a)^2 \nonumber \\
  &=& \sum_a p_a {\text Tr}(\sum_{a_i} q_{a_i} X_{a_i})^2 \nonumber \\
  & \leq &  \sum_a \sum_{a_i} p_a q_{a_i} {\text Tr} (X_{a_i})^2.  \label{D}
\end{eqnarray}
Here convexity of the function $f(X)= {\text Tr} (X^2)$ has been used.
Now let $P$ denote the projection operator into $\Lambda$,
the space spanned by the $d$ eigenvectors corresponding to
the $d$ largest eigenvalues of $\rho$.
By assumption,
\begin{equation}
{\text Tr} (\rho P) > 1  - \xi . \label{xi}
\end{equation}
Consider
\begin{eqnarray}
& & \sum_a \sum_{a_i} p_a  q_{a_i} {\text Tr} [ \Pi_{a_i} (1-P)] \nonumber \\
&=& \sum_a p_a {\text Tr} [ \sum_{a_i} q_{a_i} \Pi_{a_i} (1-P)] \nonumber \\
&=& \sum_a p_a {\text Tr} [ \Pi_a (1-P)] \nonumber \\
&=&  {\text Tr} [ \sum_a p_a \Pi_a (1-P)]\nonumber \\
&=& {\text Tr} [ \rho (1-P)]\nonumber \\
&<& \xi . \label{mtp}
\end{eqnarray}
Notice that with Eqs.\ (\ref{D}) and (\ref{mtp})
we have essentially reduced the case of mixed
signal states
to that of pure signal states with a priori probabilities $p_a q_{a_i}$.
In what follows, we shall therefore consider the case of pure signal states
only. For simplicity, we also suppress the index $a$.
Write $| a_i \rangle $ in terms of its components in $\Lambda$ and
$\Lambda^\perp$:
\begin{equation}
| a_i \rangle = \alpha_i |l_i \rangle + \beta_i | m_i \rangle
\label{component}
\end{equation}
where $\alpha_i , \beta_i \geq 0$ , $\alpha^2 + \beta^2 =1$,
$|l_i \rangle \in \Lambda $ and
$ | m_i \rangle \in \Lambda^\perp$.
For $\Pi_i =| a_i \rangle \langle a_i |$, we have
\begin{eqnarray}
\Pi_i &= & \alpha_i^2  |l_i \rangle \langle l_i |
 + \alpha_i \beta_i |l_i  \rangle \langle m_i | \nonumber \\
 & & +\alpha_i \beta_i | m_i \rangle \langle l_i |
 + \beta_i^2 | m_i \rangle \langle m_i |.
\label{acomponent}
\end{eqnarray}
$\Pi_i$ is associated to
\begin{equation}
W_i = \alpha_i^2  |l_i \rangle \langle l_i |
 + \beta_i^2 | 0 \rangle \langle 0 |.
\label{wcomponent}
\end{equation}
It is then a simple exercise to check that
\begin{eqnarray}
\sum_i p_i {\text Tr} (\Pi_i - W_i)^2  &=&  2 \sum_i p_i \beta_i^2 \nonumber \\
         &=& 2 \sum_i {\text Tr} [ \Pi_i (1-P) ] < 2 \xi  .
\label{wacomponent}
\end{eqnarray}
This completes our proof of Lemma 1.

{\em Lemma 2.} Consider any coding scheme
\begin{equation}
\Pi_i \longleftrightarrow
W_i ~~~i=1, \cdots , m  \label{lemma1}
\end{equation}
where $W_i$ is a density matrix supported on some $d$-dimensional
subspace $D$ of ${\cal H}_n$. If the sum of the $d$ largest eigenvalues
of $\rho$ is $\eta$, then the distortion $D \geq \sum_i p_i {\text Tr} \Pi_i^2
-2 \eta$.

{\em Proof.} Let us denote the projection into $D$ by $P'$ and
the projection into the space spanned by the $d$ eigenvectors with
the $d$ largest eigenvalues by $P$.
By assumption, $\sum_i p_i {\text Tr} [\Pi_i P'] \leq {\text Tr} [\rho  P]
= \eta .$
\begin{eqnarray}
D &=& \sum_i p_i {\text Tr} [ \Pi_i - W_i]^2 \nonumber \\
  &\geq& \sum_i p_i {\text Tr} \Pi_i^2
-2 \sum_i p_i {\text Tr} [\Pi_i W_i] \nonumber \\
  &\geq& \sum_i p_i {\text Tr} \Pi_i^2
-2 \sum_i p_i {\text Tr} [\Pi_i P'] \nonumber \\
  &\geq&  \sum_i p_i {\text Tr} \Pi_i^2
-2 \eta  .  \label{lemma2}
\end{eqnarray}

Having proved the two lemmas, we proceed to prove the main theorem.
For this, we make use of the ``Asymptotic Equipartition Property(AEP)''
(an analog
of the weak law of large numbers) in classical information theory. The
weak law of large numbers states that for independent, identically
distributed (i.i.d.) random variables, ${1 \over N} \sum_{i=1}^N X_i$ is
close to its expected value $E(X)$ for a large $N$. Functions of independent
random variables are also independent random variables. Since $X_i$'s are
i.i.d, so are $\log_2 p(X_i)$'s. Applying the weak law of large number to
$\log_2 p(X_i)$'s, we obtain the AEP, which states that
${1  \over N} \log_2 {1 \over p(X_1,X_2, \cdots , X_N)}$ is close to the
entropy $H$. Here $X_1, X_2,  \cdots , X_N$ are i.i.d. random variables
and $p(X_1,X_2, \cdots , X_N) = p(X_1) p(X_2) \cdots p(X_N)$ is the
probability of the occurrence of the sequence $X_1, X_2,  \cdots , X_N$.
Therefore, it is highly likely that the probability assigned
to an observed sequence
is close to $2^{-NH}$.

This enables us to divide the set of all possible sequences into
two subsets, the set of ``typical sequences'' , where the sample
entropy is close to the true entropy, and the atypical set,
which contains all other sequences. In classical noiseless
coding theorem, we just choose our codewords in one-one correspondence
with the typical set. In other words, we only code all the typical
sequence. If an atypical sequence occurs, we accept failure. The important
point is that the probability for a sequence to be in the atypical set
is small as $N$ gets large.

{\em Proof of the quantum noiseless theorem for mixed states.}
(a) Let $\lambda_1 ,\lambda_2, \cdots,\lambda_n$ be the eigenvalues of
the density matrix $\rho$ of a quantum source $A$. Consider
$\lambda_1 ,\lambda_2, \cdots,\lambda_n$ as the probabilities of a
probability distribution $\cal P$. The Shannon entropy
$H({\cal P})$ is the same as the von Neumann entropy $S(\rho)$.
Note also that the $K$-blocked version $A_K$ of $A$ has a density matrix
$\rho_K = \bigotimes^K \rho$. The AEP states that for
sufficiently large $K$, there exists a set of
$2^{K(S+ \delta)}$ eigenvalues of $\rho_K$ with a sum of eigenvalues
greater than $1 - \epsilon/2$. Therefore, the sum of the
$2^{K(S+ \delta)}$ {\em largest}
eigenvalues must be larger than $1 - \epsilon/2$.
By Lemma 1, there exists a coding scheme for $A_K$ which uses
$K(S +\delta)$ qubits
per signal for $A_K$ and
has distortion $D < \epsilon$.

(b) Using the weak law of large numbers, it can be shown that, for
all sufficiently large $K$,
any subset of ${\cal P}_K$ of size less than $2^{K(S- \delta)}$ has probability
less than $\epsilon$. (See Ref.\ \cite{JozSch}.)
In particular, the sum of the  $2^{K(S- \delta)}$
largest eigenvalues will still be less than $\epsilon$.
By Lemma 2, we find that for all sufficiently large $K$, any
coding scheme with $K(S- \delta)$ qubits per signal will have
distortion $D \geq \sum_i p_i {\text Tr} \Pi_i^2
-2 \eta$.

This completes our proof of the noiseless coding theorem for mixed states.

Note that this theorem applies only to approximate transposition
coding schemes.
Is it possible to devise a more efficient coding scheme?
The anwer is yes \cite{comm}.
Mixed state signals might be re-constituted from a compressed version
by adjoining an ancilla in a standard state, and applying a measurement
process. Suppose, for instance, that there are two signals
$\rho_1$ and $\rho_2$ with probabilities
$p_1$ and $p_2$ respectively, and that these signals live in a 4-dimensional
space
with supports in two 2-dimensional
subspaces, which are {\em orthogonal} to each other.
We can compress the data as follows:
{\em Measure} the signal.
Since the two signals have orthogonal supports,
the measurement tells us
with certainty which of the two signals we are given.
Record the possible outcomes of our measurement by
pure orthogonal (i.e. classical)
states
$|1 \rangle$ and $|2 \rangle$ occurring with probabilities $p_1$
and $p_2 $.
It follows from the Shannon's classical noiseless coding
theorem that the signal can further be compressed to
the Shannon entropy $H(p_1, p_2)$ qubits/signal.
To reconstitute the signals, we simply decode (and decompose)
the classical signal
and represent each of the binary digit $0$ or $1$ in the resulting sequence
by a density matrice $\rho_1$ or $\rho_2$ accordingly.
But $H(p_1,p_2)$ is less than   $S(p_1 \rho_1 + p_2 \rho_2)$
(if either $\rho_1$ or $\rho_2$ is a mixed state).
The limit set by the mixed state coding theorem
is, thus, beaten by the above method.
A natural
question to ask is: what is the information theoretic limit of the compression
rate of mixed-state signals that {\em no} coding scheme can surpass?

Another point to
note is that Shannon's more important results deal with channels with noise.
The information
capacity of
a noisy channel deserves further investigations.

After the completion of this
work we learned that Jozsa\cite{Jozsa}
has proven essentially the same result, using Uhlmann's
transition probability formula\cite{Uhl} as a fidelity function\cite{fide}.
We thank R. Jozsa for bringing his unpublished results to our attention.
Helpful discussions with H. F. Chau,
K. Y. Szeto and F. Wilczek are also gratefully acknowledged.
This work was supported in part by DOE DE-FG02-90ER40542 and
HKTIIT 92/93.002 .

\end{document}